\input harvmac

\def\quad{{\ \ }}

\def\ep{\epsilon}

\let\includefigures=\iftrue
\newfam\black
\includefigures
\input epsf
\def\figin{\epsfcheck\figin}\def\figins{\epsfcheck\figins}
\def\epsfcheck{\ifx\epsfbox\UnDeFiNeD
\message{(NO epsf.tex, FIGURES WILL BE IGNORED)}
\gdef\figin##1{\vskip2in}\gdef\figins##1{\hskip.5in}
\else\message{(FIGURES WILL BE INCLUDED)}%
\gdef\figin##1{##1}\gdef\figins##1{##1}\fi}
\def\DefWarn#1{}
\def\figinsert{\goodbreak\midinsert}
\def\ifig#1#2#3{\DefWarn#1\xdef#1{fig.~\the\figno}
\writedef{#1\leftbracket fig.\noexpand~\the\figno}%
\figinsert\figin{\centerline{#3}}\medskip\centerline{\vbox{\baselineskip12pt
\advance\hsize by -1truein\noindent\footnotefont{\bf Fig.~\the\figno:}
#2}}
\bigskip\endinsert\global\advance\figno by1}
\else
\def\ifig#1#2#3{\xdef#1{fig.~\the\figno}
\writedef{#1\leftbracket fig.\noexpand~\the\figno}%
#2}}
\global\advance\figno by1}
\fi


\def\sym{ \> {\vcenter {\vbox
{\hrule height.6pt
\hbox {\vrule width.6pt height5pt
\kern5pt
\vrule width.6pt height5pt
\kern5pt
\vrule width.6pt height5pt}
\hrule height.6pt}
}
} \>
}
\def\fund{ \> {\vcenter {\vbox
{\hrule height.6pt
\hbox {\vrule width.6pt height5pt
\kern5pt
\vrule width.6pt height5pt }
\hrule height.6pt}
}
} \>
}
\def\anti{ \> {\vcenter {\vbox
{\hrule height.6pt
\hbox {\vrule width.6pt height5pt
\kern5pt
\vrule width.6pt height5pt }
\hrule height.6pt
\hbox {\vrule width.6pt height5pt
\kern5pt
\vrule width.6pt height5pt }
\hrule height.6pt}
}
} \>
}


\lref\BaggerJR{
  J.~Bagger and N.~Lambert,
  ``Gauge Symmetry and Supersymmetry of Multiple M2-Branes,''
  arXiv:0711.0955 [hep-th].
}

\lref\BerensteinJQ{
  D.~E.~Berenstein, J.~M.~Maldacena and H.~S.~Nastase,
  JHEP {\bf 0204}, 013 (2002)
  [arXiv:hep-th/0202021].
}

\lref\VanProeyenNI{
  A.~Van Proeyen,
  ``Tools for supersymmetry,''
  arXiv:hep-th/9910030.
}

\lref\KalloshYU{
  R.~Kallosh and D.~Sorokin,
  ``Dirac action on M5 and M2 branes with bulk fluxes,''
  JHEP {\bf 0505}, 005 (2005)
  [arXiv:hep-th/0501081].
}

\lref\ChenIR{
B.~Chen, W.~He, J.~B.~Wu and L.~Zhang,
``M5-branes and Wilson Surfaces,''
JHEP {\bf 0708}, 067 (2007)
[arXiv:0707.3978 [hep-th]].
}

\Title{
\vbox{\baselineskip12pt
\hbox{arXiv:0804.2186}}}
{\vbox{\centerline{Matrix Theory of  Type IIB Plane Wave from Membranes}}}

\smallskip
\smallskip

\vskip-5pt
\centerline{Jaume Gomis$^{1a}$, Amir Jafari Salim$^{1,2b}$ and 
Filippo Passerini$^{1,2c}$}
\smallskip

\smallskip
\smallskip
\bigskip
\medskip
\centerline{\it Perimeter Institute for Theoretical Physics}
\centerline{\it Waterloo, Ontario N2L 2Y5, Canada$^1$}
\bigskip
\bigskip
\centerline{\it Department of Physics and Astronomy}
\centerline{\it University of Waterloo,  Ontario N2L 3G1, Canada$^2$}
\vskip .1in

\footnote{${}^{}$}{${}^{a}$\tt jgomis@perimeterinstitute.ca}
\footnote{${}^{}$}{${}^{b}$\tt ajafarisalim@perimeterinstitute.ca}
\footnote{${}^{}$}{${}^{c}$\tt fpasserini@perimeterinstitute.ca}

\vskip .2in
\centerline{\bf{Abstract}}
\medskip
\medskip

We write down a maximally supersymmetric one parameter deformation of the field theory action of Bagger and Lambert. We show that this theory on $R\times T^2$ is invariant under the superalgebra of the maximally supersymmetric Type IIB plane wave. It is  argued that this theory holographically describes the Type IIB plane wave in the discrete light-cone quantization (DLCQ).

\Date{04/2008}

\newsec{Introduction and Discussion}

\lref\DuffXZ{
  M.~J.~Duff and K.~S.~Stelle,
  ``Multi-membrane solutions of D = 11 supergravity,''
  Phys.\ Lett.\  B {\bf 253}, 113 (1991).
}
\lref\BergshoeffQX{
  E.~Bergshoeff, E.~Sezgin and P.~K.~Townsend,
  ``Properties of the Eleven-Dimensional Super Membrane Theory,''
  Annals Phys.\  {\bf 185}, 330 (1988).
}
\lref\BandresVF{
  M.~A.~Bandres, A.~E.~Lipstein and J.~H.~Schwarz,
  ``N = 8 Superconformal Chern--Simons Theories,''
  arXiv:0803.3242 [hep-th].
}
\lref\SchwarzYJ{
  J.~H.~Schwarz,
  ``Superconformal Chern-Simons theories,''
  JHEP {\bf 0411}, 078 (2004)
  [arXiv:hep-th/0411077].
}
\lref\AharonyTH{
  O.~Aharony, M.~Berkooz, S.~Kachru, N.~Seiberg and E.~Silverstein,
  ``Matrix description of interacting theories in six dimensions,''
  Adv.\ Theor.\ Math.\ Phys.\  {\bf 1}, 148 (1998)
  [arXiv:hep-th/9707079].
}

\lref\BasuED{
  A.~Basu and J.~A.~Harvey,
  ``The M2-M5 brane system and a generalized Nahm's equation,''
  Nucl.\ Phys.\  B {\bf 713}, 136 (2005)
  [arXiv:hep-th/0412310].
}

\lref\BaggerSK{
  J.~Bagger and N.~Lambert,
  ``Modeling multiple M2's,''
  Phys.\ Rev.\  D {\bf 75}, 045020 (2007)
  [arXiv:hep-th/0611108].
}
\lref\BaggerVI{
  J.~Bagger and N.~Lambert,
  ``Comments On Multiple M2-branes,''
  JHEP {\bf 0802}, 105 (2008)
  [arXiv:0712.3738 [hep-th]].
}
\lref\GustavssonVU{
  A.~Gustavsson,
  ``Algebraic structures on parallel M2-branes,''
  arXiv:0709.1260 [hep-th].
}
\lref\GustavssonDY{
  A.~Gustavsson,
  ``Selfdual strings and loop space Nahm equations,''
  arXiv:0802.3456 [hep-th].
}

The supersymmetric worldvolume theory of a single M2-brane in an arbitrary  eleven dimensional supergravity background was found twenty years ago \BergshoeffQX. The realization that branes in eleven dimensional supergravity are related by string dualities to D-branes
\lref\PolchinskiMT{
  J.~Polchinski,
  ``Dirichlet-Branes and Ramond-Ramond Charges,''
  Phys.\ Rev.\ Lett.\  {\bf 75}, 4724 (1995)
  [arXiv:hep-th/9510017].
}
\PolchinskiMT\
 and that the low energy effective field theory of coincident D-branes is described by non-abelian  super Yang-Mills theory
 \lref\WittenIM{
  E.~Witten,
  ``Bound states of strings and p-branes,''
  Nucl.\ Phys.\  B {\bf 460}, 335 (1996)
  [arXiv:hep-th/9510135].
} \WittenIM, naturally prompts the search for the worldvolume theory of coincident M2-branes. 

Recently, Bagger and Lambert have made an explicit proposal  \BaggerJR\ for the Lagrangian description of the low energy limit of a stack of coincident M2-branes (see also the 
work of Gustavsson \GustavssonVU). This work -- which builds upon their prior paper \BaggerSK\ -- incorporates the realization in \BasuED\   that the theory on coincident M2-branes   should have generalized fuzzy-funnel configurations
\lref\ConstableAC{
  N.~R.~Constable, R.~C.~Myers and O.~Tafjord,
  ``The noncommutative bion core,''
  Phys.\ Rev.\  D {\bf 61}, 106009 (2000)
  [arXiv:hep-th/9911136].
}
\ConstableAC\ 
described by generalized Nahm equations, and  the observation in  \SchwarzYJ\ that the  putative gauge field of the theory should have a Chern-Simons like action.

In this paper we   construct a one parameter deformation of the Bagger-Lambert theory \BaggerJR\ which is maximally supersymmetric. We add to their Lagrangian a mass term for all the eight scalars and fermions\foot{The deformation of the theory on multiple $M2$-branes was first considered by Bena
\lref\BenaZB{
  I.~Bena,
  ``The M-theory dual of a 3 dimensional theory with reduced supersymmetry,''
  Phys.\ Rev.\  D {\bf 62}, 126006 (2000)
  [arXiv:hep-th/0004142].
}
\BenaZB.}
\eqn\lagpie{\eqalign{{\cal L}_{mass}= -{1\over2}\mu^2\hskip+1pt\hbox{Tr}\left(X^{I},X^{I}\right)
+{i\over 2}\mu\hskip+1pt\hbox{Tr}\left(\bar\Psi\Gamma_{3456},  \Psi\right),}}
and a Myers-like 
\lref\MyersPS{
  R.~C.~Myers,
  ``Dielectric-branes,''
  JHEP {\bf 9912}, 022 (1999)
  [arXiv:hep-th/9910053].
}
\MyersPS\  flux-inducing $SO(4)\times SO(4)$ invariant potential\foot{See also
\lref\TaylorGQ{
  W.~Taylor and M.~Van Raamsdonk,
  ``Multiple D0-branes in weakly curved backgrounds,''
  Nucl.\ Phys.\  B {\bf 558}, 63 (1999)
  [arXiv:hep-th/9904095].
}
\lref\TaylorPR{
  W.~Taylor and M.~Van Raamsdonk,
  ``Multiple Dp-branes in weak background fields,''
  Nucl.\ Phys.\  B {\bf 573}, 703 (2000)
  [arXiv:hep-th/9910052].
}
\TaylorGQ\TaylorPR.} for the scalars
\eqn\lagpieb{
{\cal L}_{flux}=-{1\over 6}\mu\hskip+1pt \varepsilon^{ABCD}\hbox{Tr}([X^A,X^B,X^C],X^D)-{1\over 6}\mu\hskip+1pt \varepsilon^{A'B'C'D'}\hbox{Tr}([X^{A'},X^{B'},X^{C'}],X^{D'})}
and show that the theory is supersymmetric. The possibility of adding the scalar mass term and the potential term for  four of the scalars was considered in \BaggerVI.
Here we show that if we give a mass to all the scalars and fermions and turn on the  potential \lagpieb\ for all the scalars that we can find a deformation of  the supersymmetry transformations of the Bagger-Lambert theory \BaggerJR\ in such a way that the deformed field theory remains fully supersymmetric. This construction yields a novel maximally supersymmetric, Poincare invariant three dimensional field theory.

We further argue that the deformed field theory  compactified on $R\times T^2$ provides the Matrix theory 
\lref\BanksVH{
  T.~Banks, W.~Fischler, S.~H.~Shenker and L.~Susskind,
  ``M theory as a matrix model: A conjecture,''
  Phys.\ Rev.\  D {\bf 55}, 5112 (1997)
  [arXiv:hep-th/9610043].
}
\BanksVH\
description\foot{In  \BerensteinJQ (see also
\lref\DasguptaHX{
  K.~Dasgupta, M.~M.~Sheikh-Jabbari and M.~Van Raamsdonk,
  ``Matrix perturbation theory for M-theory on a PP-wave,''
  JHEP {\bf 0205}, 056 (2002)
  [arXiv:hep-th/0205185].
}
\DasguptaHX), an analogous  deformation of the D0  brane Lagrangian  
was proposed as 
 the Matrix theory description of the maximally supersymmetry plane wave of eleven dimensional supergravity.}
of Type IIB string theory on the maximally supersymmetric plane wave\foot{Or given the interpretation in 
\lref\LambertET{
  N.~Lambert and D.~Tong,
  ``Membranes on an Orbifold,''
  arXiv:0804.1114 [hep-th].
}
\lref\DistlerMK{
  J.~Distler, S.~Mukhi, C.~Papageorgakis and M.~Van Raamsdonk,
  ``M2-branes on M-folds,''
  arXiv:0804.1256 [hep-th].
}
\LambertET\DistlerMK\ for the known $3$-algebra ${\cal A}_4$, 
possibly an orientifold projection of the maximally supersymmetric plane wave background.}
\lref\BlauNE{
  M.~Blau, J.~Figueroa-O'Farrill, C.~Hull and G.~Papadopoulos,
  ``A new maximally supersymmetric background of IIB superstring theory,''
  JHEP {\bf 0201}, 047 (2002)
  [arXiv:hep-th/0110242].
}
\BlauNE. We show   that the deformed field theory on $R\times T^2$ has as its algebra of symmetries the superisometry algebra of the Type IIB plane wave, as expected from a holographic dual theory. The deformed field theory on $R\times T^2$ is proposed as the   nonperturbative formulation of the Type IIB string theory in the discrete lightcone quantization (DLCQ).

We show that the supersymmetric ground states of the deformed theory are given by a discrete set of states that have an interpretation\foot{These states have yet another  space-time interpretation as M2-branes polarizing in the presence of flux into M5-branes with $S^3$ topology. 
 The supergravity description of these
ground states of the deformed theory were found in 
\lref\LinNB{
  H.~Lin, O.~Lunin and J.~M.~Maldacena,
  ``Bubbling AdS space and 1/2 BPS geometries,''
  JHEP {\bf 0410}, 025 (2004)
  [arXiv:hep-th/0409174].
}
\LinNB\ (see also
\lref\BenaJW{
  I.~Bena and N.~P.~Warner,
  ``A harmonic family of dielectric flow solutions with maximal
  supersymmetry,''
  JHEP {\bf 0412}, 021 (2004)
  [arXiv:hep-th/0406145].
}
\BenaJW).} as a collection of fuzzy $S^3$'s \BaggerVI, where 
\eqn\bpsintro{
[X^A,X^B,X^C]=-\mu\hskip+1pt \epsilon^{ABCD}X^D,\qquad X^{A'}=0}
or alternatively:
\eqn\bpsb{
[X^{A'},X^{B'},X^{C'}]=-\mu\hskip+1pt \epsilon^{A'B'C'D'}X^{D'},\qquad 
X^{A}=0.}
  We identify these states of the deformed theory  with the states in the Hilbert space of the Type IIB plane wave with zero light-cone energy, which correspond to configurations of D3-brane giant gravitons  in the Type IIB plane wave background \BerensteinJQ\ with fixed longitudinal momentum.

It would be interesting to use the deformed field theory we write down to capture stringy physics in the Type IIB plane wave. In the regime when the bulk Type IIB string theory is weakly coupled the deformed field theory can be dimensionally reduced to 1+1 dimensions 
\lref\SethiSW{
  S.~Sethi and L.~Susskind,
  ``Rotational invariance in the M(atrix) formulation of type IIB theory,''
  Phys.\ Lett.\  B {\bf 400}, 265 (1997)
  [arXiv:hep-th/9702101].
}
\lref\BanksMY{
  T.~Banks and N.~Seiberg,
  ``Strings from matrices,''
  Nucl.\ Phys.\  B {\bf 497}, 41 (1997)
  [arXiv:hep-th/9702187].
}
  and it would be interesting to extract the Type IIB lightcone string field theory interaction vertices in the plane wave background from the reduced theory.

The Lagrangian of the deformed theory is based on the same $3$-algebra structure of \BaggerJR\ (we review it in section $2$). Even though the construction of Bagger-Lambert and in this paper certainly provide new constructions of supersymmetric field theories, the precise connection with the worldvolume physics of coincident M2-branes still remains to be understood. Currently a single $3$-algebra structure is known even though several constructions have    recently been considered  
\lref\MukhiUX{
  S.~Mukhi and C.~Papageorgakis,
  ``M2 to D2,''
  arXiv:0803.3218 [hep-th].
}
\lref\BandresVF{
  M.~A.~Bandres, A.~E.~Lipstein and J.~H.~Schwarz,
  ``N = 8 Superconformal Chern--Simons Theories,''
  arXiv:0803.3242 [hep-th].
}
\lref\GranVI{
  U.~Gran, B.~E.~W.~Nilsson and C.~Petersson,
  ``On relating multiple M2 and D2-branes,''
  arXiv:0804.1784 [hep-th].
}
\MukhiUX\BandresVF, and for the known case there are subtleties identifying the M2 content of the theory as well as and the spacetime geometry in which the M2 branes are embedded 
\lref\VanRaamsdonkFT{
  M.~Van Raamsdonk,
  ``Comments on the Bagger-Lambert theory and multiple M2-branes,''
  arXiv:0803.3803 [hep-th].
}
\BaggerVI\MukhiUX\VanRaamsdonkFT\LambertET\DistlerMK.
 The possibility of relaxing the conditions on the $3$-algebra to construct new examples  have been considered
 \lref\MorozovCB{
  A.~Morozov,
  ``On the Problem of Multiple M2 Branes,''
  arXiv:0804.0913 [hep-th].
}
\MorozovCB\GranVI\ and 
the addition of a boundary  to the theory has been 
 considered in 
\lref\BermanBE{
  D.~S.~Berman, L.~C.~Tadrowski and D.~C.~Thompson,
  ``Aspects of Multiple Membranes,''
  arXiv:0803.3611 [hep-th].
}
\BermanBE. 

Establishing in more detail the M2 brane interpretation of our deformed theory is important in understanding   the deformed field theory found in this paper as the Matrix theory description of the Type IIB plane wave.

The plan of the rest of the paper is as follows. In section $2$ we quickly review the Bagger-Lambert theory \BaggerVI\ and introduce the deformation of the Lagrangian and the supersymmetry transformations  that gives rise to a new maximally supersymmetric Lagrangian in three dimensions. In section $3$ we argue that the deformed theory on $R\times T^2$ provides the Matrix theory description of the maximally supersymmetric Type IIB plane wave. We show that the theory on $R\times T^2$ has precisely the same symmetry algebra as the Type IIB plane wave and identify the supersymmetric grounds states of the deformed theory with the states in the Type IIB plane wave with zero light-cone energy, which correspond to configurations of D3-brane giant gravitons. In Appendix A we present some details of the calculation of the deformed supersymmetry transformations while in Appendix B we write down the Noether charges of the deformed field theory on $R\times T^2$ and show that they satisfy the Type IIB plane wave superalgebra.

\lref\MotlTH{
  L. Motl,
  \hskip-3pt``Proposals on nonperturbative superstring interactions,''
 \hskip-3pt  arXiv:hep-th/9701025.
}
\lref\DijkgraafVV{
  R.~Dijkgraaf, E.~P.~Verlinde and H.~L.~Verlinde,
  ``Matrix string theory,''
  Nucl.\ Phys.\  B {\bf 500}, 43 (1997)
  [arXiv:hep-th/9703030].
}
\lref\SethiSW{
  S.~Sethi and L.~Susskind,
  ``Rotational invariance in the M(atrix) formulation of type IIB theory,''
  Phys.\ Lett.\  B {\bf 400}, 265 (1997)
  [arXiv:hep-th/9702101].
}
\lref\BanksMY{
  T.~Banks and N.~Seiberg,
  ``Strings from matrices,''
  Nucl.\ Phys.\  B {\bf 497}, 41 (1997)
  [arXiv:hep-th/9702187].
}

\newsec{Deformed Supersymmetric Field Theory}

In \BaggerJR, a new  maximally supersymmetric  Lagrangian
in three dimensions has been found. The authors have proposed that this theory  describes the low energy dynamics\foot{In this $l_p\rightarrow 0$ limit, higher derivative corrections can be ignored.}
  of a stack of M2-branes. It encodes the interactions of the eight scalar fields $X^I$ transverse to the M2-branes, the worldvolume fermions $ \Psi$ and a non-propagating gauge field $A_\mu$. The Lagrangian is given by
  \BaggerJR
\eqn\lag{\eqalign{
{\cal L} &= -{1\over2}(D_\mu X^{aI})(D^\mu X^{I}_{a})
+{i\over 2}\bar\Psi^a\Gamma^\mu D_\mu \Psi_a
+{i\over4}\bar\Psi_b\Gamma_{IJ}X^I_cX^J_d\Psi_a f^{abcd}+\cr
& - V+{1\over2}\varepsilon^{\mu\nu\lambda}(f^{abcd}A_{\mu
ab}\partial_\nu A_{\lambda cd} +{2\over 3}f^{cda}{}_gf^{efgb}
A_{\mu ab}A_{\nu cd}A_{\lambda ef}),}}
where $V$ is the scalar potential 
\eqn\potential{\eqalign{V  =& {1\over 12}f^{abcd}f^{efg}{}_d
X^I_aX^J_bX^K_cX^I_eX^J_fX^K_g}\equiv{1\over2\cdot 3!}\hbox{Tr}([X^I,X^J,X^K][X^I,X^J,X^K])}
and the covariant derivative of a field $\Phi$ is given by
\eqn\covari{D_{\mu} \Phi^a = \partial_{\mu} \Phi^a -\tilde{A}^{a}_{\mu b}
 \Phi^b,}
where $\tilde{A}^{a}_{\mu b}\equiv f^a_{~bcd} A_{\mu}^{cd}$.

This theory is based on a novel algebraic structure  \BaggerJR,  a $3$-algebra ${\cal A}_n$ with generators $T^a$ -- where   $a=1,\ldots \hbox{dim}
\ {\cal A}=n$ --  and on a $3$-product:
\eqn\prod{
[T^a,T^b,T^c]=f^{abc}_{~~~~~d}T^d.}
 In \BaggerJR, the structure constants $f^{abcd}$   are taken to be totally antisymmetric and to satisfy the fundamental identity
\eqn\fun{
f^{aef}_{~~~~~g}\,f^{bcdg}-
f^{bef}_{~~~~~g}\,f^{acdg}+
f^{cef}_{~~~~~g}\,f^{abdg}-
f^{def}_{~~~~~g}\,f^{abcg}=0
}
which generalizes the familiar Jacobi identity of Lie algebras. The algebra indices are contracted with a prescribed non-degenerate metric $h^{ab}=\hbox{Tr}(T^a,T^b)$. Thus far the only examples of 3-algebras found are of the type ${\cal A}_4\oplus {\cal A}_4\oplus\ldots\oplus {\cal A}_4\oplus C_1\oplus\ldots C_l$, where ${\cal A}_4$ is defined by $f^{abcd}=\epsilon^{abcd}$ and $C_i$ denote central elements in the algebra. The supersymmetric deformation we find in this paper  applies to any 3-algebra   with totally antisymmetric structure constants  which satisfies the fundamental identity \fun.

We now find a  deformation of the action and supersymmetry 
transformations of the action of Bagger and Lambert \BaggerJR\ that is maximally supersymmetric. The new Lagrangian is given by
\eqn\newlag{\tilde{{\cal L}}={\cal L}+{\cal L}_{mass}+{\cal L}_{flux},} 
where ${\cal L}$ is the Bagger-Lambert theory in \lag\ and:
\eqn\lagpieces{\eqalign{{\cal L}_{mass}&= -{1\over2}\mu^2\hskip+1pt\hbox{Tr}(X^{I},X^{I})
+{i\over 2}\mu\hskip+1pt\hbox{Tr}(\bar\Psi\Gamma_{3456},  \Psi)\cr
{\cal L}_{flux}&=-{1\over 6}\mu\hskip+1pt \varepsilon^{ABCD}\hbox{Tr}([X^A,X^B,X^C],X^D)-{1\over 6}\mu\hskip+1pt \varepsilon^{A'B'C'D'}\hbox{Tr}([X^{A'},X^{B'},X^{C'}],X^{D'}).}}
The transverse index  has been decomposed  as  $I=(A,A')$ where $A=3,4,5,6$ and $A'=7,8,9,10$ and  $\Psi$ is an eleven dimension Majorana spinor satisfying the constraint $\Gamma_{012}\Psi=-\Psi$, where the $\Gamma$-matrices satisfy the Clifford algebra in eleven dimensions. This deformation of the Lagrangian is analogous to the deformation of  the Lagrangian of D0-branes considered in \BerensteinJQ. This deformation when restricted to only four of the scalars has been  considered in \BaggerVI.

The deformed  Lagrangian now depends on the paramater $\mu$. 
The mass term ${\cal L}_{mass}$ gives mass to all the scalars and fermions in the theory, while ${\cal L}_{flux}$ has the interpretation of the scalar potential\foot{We note that if we use the proposal made by Mukhi and Papageorgakis \MukhiUX\ to obtain by compactification the theory on D2 branes, that ${\cal L}_{flux}$ does indeed reduce to the known Myers term.} generated by a background four-form flux of eleven dimensional supergravity, of the type found by Myers \MyersPS\ (see also \TaylorGQ\TaylorPR)
in the context of $D$-branes in the presence of background fluxes.

The deformed theory \newlag\ breaks the $SO(8)$ R-symmetry of the undeformed theory \lag\ down to $SO(4)\times SO(4)$. The deformed theory is nevertheless invariant under sixteen linearly realized supersymmetries. The  supersymmetry transformations of the deformed theory are given by
\eqn\susynew{\eqalign{\tilde{\delta} X^I =&\hskip+1pt i\bar\epsilon\Gamma^I\Psi\cr
\tilde{\delta} \Psi =&\hskip+1pt D_\mu X^I\Gamma^\mu \Gamma^I\epsilon -{1\over 6}
[X^I,X^J,X^K]\Gamma^{IJK}\epsilon - \mu\hskip+1pt\Gamma_{3456}\Gamma^{I}X^I\epsilon\cr
 \tilde{\delta} \tilde{A}_{\mu}{}^b{}_a =&\hskip+1pt i\bar\epsilon
\Gamma_\mu\Gamma_IX^I_c\Psi_d f^{cdb}{}_{a},}}
where $\epsilon$ is a constant eleven dimensional Majorana spinor satisfying the constraint $\Gamma_{012}\epsilon=\epsilon$. By setting   $\mu\rightarrow 0$ we recover the  supersymmetry transformations of the undeformed theory \lag\ found by Bagger-Lambert \BaggerJR. The proof that the action \newlag\ is invariant under the supersymmetry transformations is summarized in  Appendix $A$.

The deformed action \newlag\ is also invariant under sixteen non-linearly realized supersymmetries if the 3-algebra ${\cal A}_n$ has a central element $C=T^0$, so that $f^{abc0}=0$.  Then the action \newlag\ is invariant under the following non-linear supersymmetry transformations\foot{The Bagger-Lambert theory \lag\ is also invariant under the sixteen nonlinearly realized supersymmetries obtained by setting $\mu\rightarrow 0$.}
\eqn\susynl{\eqalign{\delta_{n} X^I_a =&\hskip+1pt 0\cr
\delta_{n} \Psi =&\hskip+1pt \exp\left({-{\mu\over 3} \Gamma_{3456}\Gamma_{\mu}\sigma^{\mu}}\right) T^0\eta \cr
\delta_{n}\tilde{A}_{\mu}{}^b{}_a =&\hskip+1pt 0}}
where now $\eta$ is an eleven dimensional Majorana spinor satisfying the constraint  
$\Gamma_{012}\eta=-\eta$ and $\sigma^\mu$ are the three dimensional field theory  coordinates.

The field theory with Lagrangian \newlag\lagpieces\ and with supersymmetry transformations \susynew\susynl\ defines a novel  maximally supersymmetric Poincare invariant three dimensional field theory with $SO(4)\times SO(4)$ R-symmetry.

\newsec{Deformed Theory as DLCQ of Type IIB Plane Wave}

In 
\lref\SethiSW{
  S.~Sethi and L.~Susskind,
  ``Rotational invariance in the M(atrix) formulation of type IIB theory,''
  Phys.\ Lett.\  B {\bf 400}, 265 (1997)
  [arXiv:hep-th/9702101].
}
\lref\BanksMY{
  T.~Banks and N.~Seiberg,
  ``Strings from matrices,''
  Nucl.\ Phys.\  B {\bf 497}, 41 (1997)
  [arXiv:hep-th/9702187].
}
\SethiSW\BanksMY,  the theory of coincident M2-branes on $R\times T^2$ 
was argued\foot{At that time there was no Lagrangian description of the coincident M2-brane theory.} 
to provide the Matrix theory 
\BanksVH\ description of Type IIB string theory in flat space, extending the Matrix string theory description in 
\lref\MotlTH{
 \hskip-3ptL. Motl,
  \hskip-3pt``Proposals on nonperturbative superstring interactions,''
 \hskip-3pt  arXiv:hep-th/9701025.
}
\lref\DijkgraafVV{
  R.~Dijkgraaf, E.~P.~Verlinde and H.~L.~Verlinde,
  ``Matrix string theory,''
  Nucl.\ Phys.\  B {\bf 500}, 43 (1997)
  [arXiv:hep-th/9703030].
}
\MotlTH\DijkgraafVV\ to Type IIB string theory.  

In this section we argue that the three dimensional deformed field theory \newlag\ on $R\times T^2$ 
provides the Matrix theory\foot{For a different proposal for the Matrix theory of the Type IIB plane wave see
\lref\SheikhJabbariIK{
  M.~M.~Sheikh-Jabbari,
  ``Tiny graviton matrix theory: DLCQ of IIB plane-wave string theory, a
  conjecture,''
  JHEP {\bf 0409}, 017 (2004)
  [arXiv:hep-th/0406214].
}
\SheikhJabbariIK. For the DLCQ description of the  plane wave in terms of a sector of a quiver gauge theory see 
\lref\MukhiCK{
  S.~Mukhi, M.~Rangamani and E.~P.~Verlinde,
  ``Strings from quivers, membranes from moose,''
  JHEP {\bf 0205}, 023 (2002)
  [arXiv:hep-th/0204147].
}
\MukhiCK. See also
\lref\GopakumarDQ{
  R.~Gopakumar,
  ``String interactions in PP-waves,''
  Phys.\ Rev.\ Lett.\  {\bf 89}, 171601 (2002)
  [arXiv:hep-th/0205174].
}
\lref\BonelliMB{
  G.~Bonelli,
  ``Matrix strings in pp-wave backgrounds from deformed super Yang-Mills
  theory,''
  JHEP {\bf 0208}, 022 (2002)
  [arXiv:hep-th/0205213].
}
\lref\LozanoJR{
  Y.~Lozano and D.~Rodriguez-Gomez,
  ``Type II pp-wave matrix models from point-like gravitons,''
  JHEP {\bf 0608}, 022 (2006)
  [arXiv:hep-th/0606057].
}
\GopakumarDQ\BonelliMB\LozanoJR.}
description of the maximally supersymmetric Type IIB plane wave background
\BlauNE:
\eqn\pp{\eqalign{
&ds^2=2dx^+dx^-  -\mu^2x^Ix^Idx^+dx^++dx^Idx^I\cr
&F_{+1234}=F_{+5678}=2\mu.}}
As in the case of flat space, the modular parameter $\tau$ of the torus on which the deformed field theory is defined determines the complexified coupling constant of Type IIB string theory $\tau=C_0+i/g_s$
\lref\SchwarzDK{
  J.~H.~Schwarz,
  ``An SL(2,Z) multiplet of type IIB superstrings,''
  Phys.\ Lett.\  B {\bf 360}, 13 (1995)
  [Erratum-ibid.\  B {\bf 364}, 252 (1995)]
  [arXiv:hep-th/9508143].
}
\SchwarzDK.

\lref\Gomispp{
  J.~Gomis, A.~J.~Salim and F.~Passerini,
  arXiv:0804.2186 [hep-th].
}

In this paper we have constructed a one parameter deformation of the 
Bagger-Lambert field
theory that preserves all the thirty-two supersymmetries. 
 It is therefore  natural to propose that the deformed theory \newlag\ found in this paper is the Matrix theory description of the Type  IIB plane wave. Also as $\mu\rightarrow 0$ the plane wave background \pp\ reduces to flat space just as  the deformed field theory \newlag\  goes over to the Bagger-Lambert theory \lag, which as the candidate theory for multiple M2-branes is the Matrix theory for flat space\foot{See \LambertET\DistlerMK\ for subtleties with this interpretation.}.

 Matrix theory  describes nonperturbatively a string/M-theory background in the discrete light cone quantization (DLCQ)
\lref\SusskindCW{
  L.~Susskind,
  ``Another conjecture about M(atrix) theory,''
  arXiv:hep-th/9704080.
} \SusskindCW. In this quantization we consider a string/M-theory background with a compactified lightlike coordinate $x^-\simeq x^-+2\pi R$ in a sector with quantized longitudinal momentum $P^+=N/R$. The Matrix theory description of a string/M-theory background with some prescribed asymptopia must realize the same symmetries as those of the asymptotic background with the  lightlike identification $x^-\simeq x^-+2\pi R$.

If we consider the DLCQ of  Type IIB string theory in $R^{1,9}$, then the $ISO(1,9)$ symmetry algebra of Minkowski space is broken by the $x^-$ identification to the centrally extended Super-Galileo algebra $SGal(1,8)$ \SusskindCW, where the central extension corresponds to $P^+$.

The Type IIB plane wave background \pp\ is invariant under thirty-two supersymmetries and under a thirty-dimensional bosonic symmetry algebra \BlauNE. Unlike in flat space, the $x^-\simeq x^-+2\pi R$ does not break any of these symmetries. It is useful to gain intuition on the action of these symmetries to notice that the bosonic symmetries of the Type IIB plane wave background \pp\ can be identified
 with the centrally extended Newton-Hooke algebra\foot{
 This algebra has appeared previously in the context of non-relativistic symmetries of string theory in e.g.
 \lref\GaoSR{
  Y.~h.~Gao,
  ``Symmetries, matrices, and de Sitter gravity,''
  arXiv:hep-th/0107067.
}
 \lref\GibbonsRV{
  G.~W.~Gibbons and C.~E.~Patricot,
  ``Newton-Hooke space-times, Hpp-waves and the cosmological constant,''
  Class.\ Quant.\ Grav.\  {\bf 20}, 5225 (2003)
  [arXiv:hep-th/0308200].
}
\lref\GomisPG{
  J.~Gomis, J.~Gomis and K.~Kamimura,
  ``Non-relativistic superstrings: A new soluble sector of AdS(5) x S**5,''
  JHEP {\bf 0512}, 024 (2005)
  [arXiv:hep-th/0507036].
}
\lref\BruguesYD{
  J.~Brugues, J.~Gomis and K.~Kamimura,
  ``Newton-Hooke algebras, non-relativistic branes and generalized pp-wave
  metrics,''
  Phys.\ Rev.\  D {\bf 73}, 085011 (2006)
  [arXiv:hep-th/0603023].
}
\GaoSR\GibbonsRV\GomisPG\BruguesYD.} $NH(1,8)$. This algebra of symmetries is the non-relativistic contraction\foot{The flux in \pp\ actually breaks the $SO(8)$ rotation symmetry of the contracted algebra down to $SO(4)\times SO(4)$.} of the isometry algebra of $AdS_9$, just like the  $Gal(1,8)$ symmetry algebra of Matrix string theory in flat space arises in the non-relativistic contraction of the isometry algebra of $R^{1,8}$. As in the case of flat space, the central extension corresponds to $P^+$. Therefore the non-central generators of $NH(1,8)$    are
given by $H,P^I,K^I,J^{AB}$ and $J^{A'B'}$, which generate time translations, spatial translations, boosts and rotations respectively, 
and where the transverse index has been decomposed as $I=(A,A')$.

The deformed field theory \newlag\ is manifestly invariant under the action of $H, J^{AB}$ and $J^{A'B'}$, which correspond in the  deformed field theory \newlag\ to the Hamiltonian and the $SO(4)\times SO(4)$ R-symmetry charges of the three dimensional field theory. The non-manifest  symmetries that remain to be realized are the translations $P^I$ and boosts $K^I$. 
We now consider the following non-linear action of these generators 
  on the fields of the deformed field theory \newlag\ 
\eqn\actionofsymm{\eqalign{
\ \delta X^I&=\hskip+1pt a^J\delta^{IJ}\cos(\mu \sigma^0) T^0\cr
 P^J:~~~~~~~~ \delta \Psi&=\hskip+1pt0\cr
\delta\tilde{A}_{\mu}{}^b{}_a &=\hskip+1pt 0}}
and
\eqn\actionofsymmb{\eqalign{
  \delta X^I&=\hskip+1pt v^J\delta^{IJ}{\sin(\mu\sigma^0)\over \mu}T^0 \cr
K^J:~~~~~~~~  \delta \Psi&=\hskip+1pt 0\cr
\delta\tilde{A}_{\mu}{}^b{}_a &=\hskip+1pt 0,}}
where $\sigma^0$ is the field theory time coordinate and $T^0$ is a central element in the $3$-algebra ${\cal A}$. Note that  in the flat space limit $\mu\rightarrow 0$ we recover the usual Galilean transformations.
Under the action of the transformations \actionofsymm\ and \actionofsymmb\ the deformed Lagrangian \newlag\ changes by a total derivative. This provides the field theory explanation for  the existence of the central extension $P^+$ in $NH(1,8)$, as central extensions of symmetry algebras are always associated with symmetry transformations that result in quasi-invariant Lagrangians.
The central extension appears in the commutator of translations and boosts:
\eqn\commcent{
[P^I,K^J]=i\delta^{IJ} P^+.}

 The original three dimensional Poincare symmetry of the field theory is broken by compactification to $R\times T^2$ to just the translation algebra. The time translation generator $H$ is identified with the Type IIB Hamiltonian. The translation generators along the $T^2$ can be identified with central charges of the superalgebra
\lref\BanksNN{
  T.~Banks, N.~Seiberg and S.~H.~Shenker,
  ``Branes from matrices,''
  Nucl.\ Phys.\  B {\bf 490}, 91 (1997)
  [arXiv:hep-th/9612157].
}
\BanksNN\BanksMY. These central charges are associated with fundamental strings and D1 strings wrapping the longitudinal direction
of the Type IIB plane wave \pp. The geometrical action of $SL(2,Z)$ on the $T^2$ on which the theory is defined exchanges the fundamental and D1 strings in the way expected from duality \SchwarzDK.

The supercharge generating the supersymmetry transformations \susynew\ correspond to the dynamical supersymmetries of  the Type IIB plane wave \pp\ while the supercharges generating the supersymmetry transformations \susynl\ correspond to the kinematical supersymmetries of the plane wave. Thus combining the bosonic  symmetries  with the supersymmetry transformations found in the section $2$ we conclude that the deformed field theory \newlag\ is invariant under $SNH(1,8)$, or equivalenty under the superisometry algebra of the Type IIB plane background \pp\ in the DLCQ.   In Appendix $B$ we write the Noether charges of the deformed field theory on $R\times T^2$ and show that the commutation relations are those of the Type IIB plane wave \pp.

Therefore the deformed field theory \newlag\ has the necessary ingredients to be the Matrix theory description of the Type IIB plane wave \pp.
 
 \subsec{Deformed Field Theory Vacua and Type IIB Plane Wave Giant Gravitons}
 
 Type IIB string theory on the plane wave background \pp\ contains in its Hilbert space states with zero light-cone energy -- where $H=0$ -- that preserve half of the supersymmetry \BerensteinJQ. They correspond to configurations of giant gravitons. A giant graviton in \pp\ is a $D3$ brane  which wraps $S^3$ or $\tilde{S}^3$  at $x^-=0$, where $S^3$  ($\tilde{S}^3$) is the sphere 
 of the first (second) $R^4$ in the plane wave geometry \pp. The radius of the giant graviton is determined by the longitudinal momentum $P^+$ carried by the $D3$-brane \BerensteinJQ:
 \eqn\radius{
 {L^2\over \alpha^\prime}=2\pi g_s\mu P^+\alpha^\prime.}
 
 When considering the DLCQ of the Type IIB plane wave, the total longitudinal momentum is quantized $P^+=N/R$. Therefore, the $H=0$ states of the DLCQ of the plane wave are labeled by partitions of $N$, and each state describes a configuration of $D3$-branes whose total longitudinal momentum is $P^+=N/R$. These $D3$-brane configurations preserve half of the supersymmetries. More precisely,  they preserve all the sixteen linearly realized supersymmetries  while they break all of the non-linearly  supersymmetries  of the plane wave background.

 The deformed field theory \newlag\ also contains in its Hilbert space zero energy states that preserve half of the supersymmetries of the theory. These ground states are described by constant 
scalar fields satisfying 
\eqn\bps{
[X^A,X^B,X^C]=-\mu\hskip+1pt \epsilon^{ABCD}X^D,\qquad X^{A'}=0}
or alternatively:
\eqn\bpsb{
[X^{A'},X^{B'},X^{C'}]=-\mu\hskip+1pt \epsilon^{A'B'C'D'}X^{D'},\qquad 
X^{A}=0,}
where we have split the transverse index $I=(A,A')$, with $A=3,4,5,6$ and $A'=7,8,9,10$.
These solutions automatically satisfy the supersymmetry
 condition\foot{The supersymmetry conditions of \SheikhJabbariIK\ were analyzed in 
\lref\SheikhJabbariMF{
  M.~M.~Sheikh-Jabbari and M.~Torabian,
  ``Classification of all 1/2 BPS solutions of the tiny graviton matrix
  theory,''
  JHEP {\bf 0504}, 001 (2005)
  [arXiv:hep-th/0501001].
}
\SheikhJabbariMF.}  $\tilde{\delta} \Psi=0$ in \susynew\ and preserve all the linearly realized supersymmetries while they break the non-linearly realized supersymmetries, just like the giant gravitons in the Type IIB plane wave \pp. It is straightforward to show that these states also have $H=0$.
 
 We identify these states of the deformed field theory with the giant graviton configurations of the Type IIB plane wave. Further work on $3$-algebras and their representation theory is important to further understand the Matrix theory proposal of this paper.

\bigbreak\bigskip\bigskip\centerline{{\bf Acknowledgements}}\nobreak

We thank  Joaquim Gomis,  Jorge Russo
and Mark Van Raamsdonk for fruitful discussions. J.G. would like to thank the University of Barcelona 
for hospitality during the final stages of this work. This research was supported by Perimeter Institute for Theoretical
Physics.  Research at Perimeter Institute is supported by the Government
of Canada through Industry Canada and by the Province of Ontario through
the Ministry of Research and Innovation. J.G.  also acknowledges further  support by an NSERC Discovery Grant.



\medskip\medskip\medskip\medskip


\appendix{A}{Supersymmetry of Deformed Field Theory}


We first note that the susy variation \susynew\ can be decomposed as 
\eqn\susydec{\tilde{\delta}=\delta_{\epsilon}+\delta_{\mu},} where  $\delta_{\epsilon}$ are given in 
\eqn\susyb{\eqalign{\delta_\epsilon X^I_a =&\hskip+1pt i\bar\epsilon\Gamma^I\Psi_a\cr
\delta_\epsilon \Psi_a =&\hskip+1pt D_\mu X^I_a\Gamma^\mu \Gamma^I\epsilon -{1\over 6}
X^I_bX^J_cX^K_d f^{bcd}{}_{a}\Gamma^{IJK}\epsilon \cr
 \delta_\epsilon\tilde{A}_{\mu}{}^b{}_a =&\hskip+1pt i\bar\epsilon
\Gamma_\mu\Gamma_IX^I_c\Psi_d f^{cdb}{}_{a}.}}
 and 
\eqn\susym{\eqalign{\delta_\mu X^I_a =&\hskip+1pt 0\cr
\delta_\mu \Psi =&\hskip+1pt -\mu \Gamma_{3456}\Gamma^{I}X^I\epsilon \cr
 \delta_\mu\tilde{A}_{\mu}{}^b{}_a =&\hskip+1pt0},}
 where $\epsilon$ is an eleven dimensional Majorana spinor subject to the constraint $\Gamma_{012}\epsilon=\epsilon$.
 Since $\tilde{{\cal L}}={\cal L} +{\cal L}_{mass}+{\cal L}_{flux}$, we have that:
 \eqn\varpieces{\tilde{\delta}\tilde{{\cal L}}=\delta_{\epsilon}{\cal L}+\delta_{\epsilon}{\cal L}_{mass}+\delta_{\epsilon}{\cal L}_{flux}+\delta_{\mu}{\cal L}+\delta_{\mu}{\cal L}_{mass}+\delta_{\mu}{\cal L}_{flux}.}
In \BaggerJR\ it has already been shown that $\delta_{\epsilon}{\cal L}=0$ up to total derivatives.  It is trivial to see that  $\delta_{\mu}{\cal L}_{flux}=0$. The other terms are: 
\eqn\demass{\eqalign{\delta_{\epsilon}{\cal L}_{mass}=& -\mu^2\hbox{Tr}(X^I,i\bar{\epsilon}\Gamma^I\Psi)+i\mu\hbox{Tr}(D_{\mu}X^I,\bar{\Psi}\Gamma_{3456}\Gamma^\mu\Gamma^{I}\epsilon)\cr & -i{1\over 6}\mu\hbox{Tr}([X^I,X^J,X^K],\bar{\Psi}\Gamma_{3456}\Gamma^{IJK}\epsilon)}}
\eqn\deramo{\eqalign{\delta_{\epsilon}{\cal L}_{flux}=&i{2\over 3}
\mu\varepsilon^{ABCD}\hbox{Tr}([X^A,X^B,X^C],\bar{\Psi}\Gamma^D\epsilon)\cr &+i{2\over 3}\mu\varepsilon^{A'B'C'D'}\hbox{Tr}([X^{A'},X^{B'},X^{C'}],\bar{\Psi}\Gamma^{D'}\epsilon)\cr=&-i{2\over 3}\mu\hbox{Tr}([X^A,X^B,X^C],\bar{\Psi}\Gamma^{ABC}\Gamma_{3456}\epsilon)\cr &+i{2\over 3}\mu\hbox{Tr}([X^{A'},X^{B'},X^{C'}],\bar{\Psi}\Gamma^{A'B'C'}\Gamma_{3456}\epsilon)
}}
In the last step of \deramo\ we have used 
\eqn\ggg{\varepsilon^{ABCD}\Gamma^D=-\Gamma^{ABC}\Gamma_{3456},\qquad \varepsilon^{A'B'C'D'}\Gamma^{D'}=-\Gamma^{A'B'C'}\Gamma_{789(10)},} and
\eqn\ep{\Gamma_{789(10)}\epsilon=-\Gamma_{3456}\epsilon,}
which  is implied by $\Gamma_{012}\epsilon=\epsilon$ and $\Gamma_{0123456789(10)}=-1$. We also have that
\eqn\dml{\eqalign{\delta_{\mu}{\cal L}=&-{i\over 2}\partial_\mu \hbox{Tr}(\bar{\Psi}\Gamma^\mu,\delta_\mu\Psi)-i\mu\hbox{Tr}(D_{\mu}X^I,\bar{\Psi}\Gamma_{3456}\Gamma^\mu\Gamma^{I}\epsilon)\cr&-i{1\over 2}\mu\hbox{Tr}([X^I,X^J,X^K],\bar{\Psi}\Gamma^{IJ}\Gamma_{3456}\Gamma^K\epsilon)}}
and that
\eqn\dmm{\delta_{\mu}{\cal L}_{mass}=\mu^2\hbox{Tr}(i\bar{\epsilon}\Gamma^I\Psi, X^I).}

Combining all the pieces together we get
\eqn\vartot{\eqalign{\tilde{\delta}\tilde{{\cal L}}=& -i{1\over 6}\mu\hbox{Tr}([X^I,X^J,X^K],\bar{\Psi}\Gamma_{3456}\Gamma^{IJK}\epsilon)\cr&-i{1\over 2}\mu\hbox{Tr}([X^I,X^J,X^K],\bar{\Psi}\Gamma^{IJ}\Gamma_{3456}\Gamma^K\epsilon)\cr&-i{2\over 3}\mu\hbox{Tr}([X^A,X^B,X^C],\bar{\Psi}\Gamma^{ABC}\Gamma_{3456}\epsilon)\cr &+i{2\over 3}\mu\hbox{Tr}([X^{A'},X^{B'},X^{C'}],\bar{\Psi}\Gamma^{A'B'C'}\Gamma_{3456}\epsilon),}} 
where we have omitted the surface term in  \dml. 
Using the identities 
\eqn\idone{\eqalign{[X^I,X^J,X^K]\Gamma_{3456}\Gamma^{IJK}=&-[X^A,X^B,X^C]\Gamma^{ABC}\Gamma_{3456}+3[X^A,X^B,X^{A'}]\Gamma^{ABA'}\Gamma_{3456}\cr -&3[X^{A'},X^{B'},X^{A}]\Gamma^{A'B'A}\Gamma_{3456}+[X^{A'},X^{B'},X^{C'}]\Gamma^{A'B'C'}\Gamma_{3456}}}
\eqn\idtwo{\eqalign{[X^I,X^J,X^K]\Gamma^{IJ}\Gamma_{3456}\Gamma^{K}=&-[X^A,X^B,X^C]\Gamma^{ABC}\Gamma_{3456}-[X^A,X^B,X^{A'}]\Gamma^{ABA'}\Gamma_{3456}\cr +&[X^{A'},X^{B'},X^{A}]\Gamma^{A'B'A}\Gamma_{3456}+[X^{A'},X^{B'},X^{C'}]\Gamma^{A'B'C'}\Gamma_{3456},}}
one can show that the right hand side of \vartot\ vanishes. This implies that the the deformed field theory is invariant under sixteen linearly realized supersymmetries.

The proposed non-linearly realized supersymmetry transformations are given by
\eqn\susynl{\eqalign{\delta_{n} X^I_a =&\hskip+1pt 0\cr
\delta_{n} \Psi =&\hskip+1pt  \exp\left({-{1\over 3}\mu \Gamma_{3456}\Gamma_{\mu}\sigma^{\mu}}\right)T^0 \eta \cr
\delta_{n}\tilde{A}_{\mu}{}^b{}_a =&\hskip+1pt 0},}
where now $\eta$ is an eleven dimensional Majorana spinor subject to the constraint $\Gamma_{012}\eta=-\eta$ and $T^0$ is a central generator of the $3$-algebra. 
The variation of the Lagrangian $\newlag$ gives
\eqn\varnonlin{\eqalign{\delta_{n}\tilde{{\cal L}}=&{i}\bar\Psi^a\Gamma^\mu (D_\mu \delta_{n}\Psi)_a
+{i\over2}\bar\Psi_b\Gamma_{IJ}X^I_cX^J_d\delta_n\Psi_a f^{abcd}+i\mu\bar{\Psi}^a\Gamma_{3456}\delta_n\Psi_a\cr &-{i\over 2}\partial_{\mu}(\bar{\Psi}^a\Gamma^\mu\delta_n\Psi_a)\cr
=&i\bar{\Psi}^0\Gamma^\mu \partial_\mu(e^{-{1\over 3}\mu \Gamma_{3456}\Gamma_{\mu}\sigma^{\mu}})\eta+i\mu\bar{\Psi}^0\Gamma_{3456} e^{-{1\over 3}\mu \Gamma_{3456}\Gamma_{\mu}\sigma^{\mu}}\eta\cr 
=&0,}}
where in the second step we used that $f^{cd0}{}_b=0$ -- $T^0$ being central -- and have ignored a total derivative. Therefore the deformed field theory \newlag\ is also invariant under sixteen non-linearly realized supersymmetries.

When the deformed field theory is placed on $R\times T^2$ the three dimensional Poincare symmetry is broken. In this case the theory is invariant under the following transformations:
\eqn\susynltorus{\eqalign{\delta_{n} X^I_a =&\hskip+1pt 0\cr
\delta_{n} \Psi =&\hskip+1pt  \exp\left({-\mu \Gamma_{3456}\Gamma_{0}\sigma^{0}}\right)T^0 \eta \cr
\delta_{n}\tilde{A}_{\mu}{}^b{}_a =&\hskip+1pt 0}.}

\appendix{B}{Noether Charges and Supersymmetry Algebra}

The  charges that generate the symmetry transformations of the deformed field theory on $R\times T^2$ are given by
\eqn\generators{\eqalign{P^+=&\int d^2 \sigma \cr
P^I=&\hskip+1pt \int d^2 \sigma\left(\Pi_0^I\cos(\mu\sigma^0)+\mu X_0^I\sin(\mu\sigma^0)\right)\cr
K^I=&\hskip+1pt \int d^2 \sigma\left(\Pi_0^I{\sin(\mu\sigma^0)\over \mu}-X^I_0\cos(\mu\sigma^0)\right)\cr
J^{AB}=&\hskip+1pt-i\int d^2 \sigma\left(\hbox{Tr}(X^{A},\Pi^{B})-\hbox{Tr}(X^{B},\Pi^{A})+{i\over 4}\hbox{Tr}(\bar{\Psi},\Gamma^{AB}\Gamma^0\Psi)\right)\cr
J^{A'B'}=&\hskip+1pt-i\int d^2 \sigma\left(\hbox{Tr}(X^{A'},\Pi^{B'})-\hbox{Tr}(X^{B'},\Pi^{A'})+{i\over 4}\hbox{Tr}(\bar{\Psi},\Gamma^{A'B'}\Gamma^0\Psi)\right)\cr
Q=&\hskip+1pt\int d^2 \sigma\Big(-\hbox{Tr}(D_\mu X^I,\Gamma^\mu\Gamma^I\Gamma^0 \Psi)-{1\over 6}\hbox{Tr}( [X^I,X^J,X^K], \Gamma^{IJK}\Gamma^0\Psi)\cr &\hskip+45pt+\mu \Gamma^{I} \Gamma_{3456}\Gamma^0\hbox{Tr}(X^I,\Psi)\Big)\cr
q=&\hskip+1pt -i\int d^2 \sigma\Gamma^0\exp\left({-\mu \Gamma_{3456}\Gamma_{0}\sigma^{0}}\right)\Psi_0,}} 
where $\int d^2 \sigma$ is the integral over the $T^2$.
The Hamiltonian of the theory is given by:
\eqn\hamil{\eqalign{H&= \Pi_a^If^{cdba}A_{0cd}X_b^I+{1\over2 }\Pi_a^I\Pi^{aI}+{1\over2 }D_iX_a^ID_iX^{aI}\cr&
+{i\over 2}\bar\Psi^a\Gamma^0  \dot{\Psi}_a-{i\over 2}\bar\Psi^a\Gamma^0 D_0 \Psi_a-{i\over 2}\bar\Psi^a\Gamma^i D_i \Psi_a\cr&
+{i\over4}\hbox{Tr}([\bar\Psi\Gamma_{IJ},\Psi,  X^I],X^J)+V+{1\over2}\mu^2\hbox{Tr}(X^{I},X^{I})-{i\over 2}\mu\hbox{Tr}(\bar\Psi,\Gamma_{3456}  \Psi)\cr
& +{1\over 6}\mu \varepsilon^{ABCD}\hbox{Tr}([X^A,X^B,X^C],X^D)+{1\over 6}\mu \varepsilon^{A'B'C'D'}\hbox{Tr}([X^{A'},X^{B'},X^{C'}],X^{D'})\cr&
+\Lambda^{cd\lambda}\dot{A}_{cd\lambda}-{1\over2}\varepsilon^{\mu\nu\lambda}(f^{abcd}A_{\mu
ab}\partial_\nu A_{\lambda cd} +{2\over 3}f^{cda}{}_gf^{efgb}
A_{\mu ab}A_{\nu cd}A_{\lambda ef}).}}
Alternatively, one can write:
\eqn\hamil{\eqalign{H&= \Pi_a^If^{cdba}A_{0cd}X_b^I+{1\over2 }\Pi_a^I\Pi^{aI}+{1\over2 }D_iX_a^ID_iX^{aI}\cr&
+{i\over 2}\bar\Psi_a\Gamma^0f^{cdba}A_{0cd}\Psi_b-{i\over 2}\bar\Psi^a\Gamma^i D_i \Psi_a\cr&
+{i\over4}\hbox{Tr}([\bar\Psi\Gamma_{IJ},\Psi,  X^I],X^J)+V+{1\over2}\mu^2\hbox{Tr}(X^{I},X^{I})-{i\over 2}\mu\hbox{Tr}(\bar\Psi,\Gamma_{3456}  \Psi)\cr
& +{1\over 6}\mu \varepsilon^{ABCD}\hbox{Tr}([X^A,X^B,X^C],X^D)+{1\over 6}\mu \varepsilon^{A'B'C'D'}\hbox{Tr}([X^{A'},X^{B'},X^{C'}],X^{D'})\cr&
-{1\over2}\varepsilon^{\mu i \lambda}(f^{abcd}A_{\mu
ab}\partial_i A_{\lambda cd}) -{1\over 3}\epsilon^{\mu\nu\lambda}f^{cda}{}_gf^{efgb}
A_{\mu ab}A_{\nu cd}A_{\lambda ef}.}}
where $i=1,2$. 

In order to calculate the algebra generated by these charges we need the canonical momenta. $\Pi^I$ is the momentum density conjugate to $X^I$ and satisfies 
\eqn\cano{[X_a^I(\sigma^i),\Pi_b^J(\sigma'{}^{i})]=i\delta^2(\sigma^i-\sigma'{}^{i})\delta_{ab}\delta^{IJ},}      
where $i=1, 2$ are the spatial coordinates on the membrane and $\Pi_a^I=D_0X_a^I$.  For the canonical commutation relation for the spinors, one must use Dirac brackets, which for the case of Majorana spinors results in the following commutation relation:
\eqn\commuferm{
\{\Psi_a^\alpha(\sigma^i),\Psi_b^\beta(\sigma'{}^i)\}=-\delta^2(\sigma^i-\sigma'{}^i)\delta_{ab}\delta^{\alpha\beta}}
where $\alpha,\beta$ are eleven dimensional spinor indices.

To compute the action of the symmetries on the fields, we compute the commutator of the charges with the fields. We get
\eqn\genercomm{\eqalign{
[P^I,X^J]=&\hskip+1pt-i\delta^{IJ}\cos(\mu\sigma^0)T^0\cr
[K^I,X^J]=&\hskip+1pt-i\delta^{IJ}{\sin(\mu\sigma^0)\over \mu}T^0\cr
[J^{AB},X^{C}]=&\hskip+1pt-X^A\delta^{BC}+X^B\delta^{AC}\cr
[J^{AB},\Psi]=&\hskip+1pt -{1\over 2} \Gamma^{AB}\Psi\cr
[J^{A'B'},X^{C'}]=&\hskip+1pt-X^{A'}\delta^{B'C'}+X^{B'}\delta^{A'C'}\cr
[J^{A'B'},\Psi]=&\hskip+1pt -{1\over 2} \Gamma^{A'B'}\Psi\cr
[\bar{\epsilon}Q,X^J]=&\hskip+1pti\bar{\epsilon}\Gamma^I\Psi\cr
[\bar{\epsilon}Q,\Psi]=&\hskip+1pt D_\mu X^I\Gamma^\mu \Gamma^I\epsilon -{1\over 6}
[X^I,X^J,X^K]\Gamma^{IJK}\epsilon - \mu \Gamma_{3456}\Gamma^{I}X^I\epsilon\cr
[\bar{\epsilon}Q, A_{ab  i}]&=\hskip+1pt i\bar{\epsilon}\Gamma_i\Gamma^IX_{[a}^I\Psi_{b]}\cr
[\bar{\eta}q,\Psi]=&\hskip+1pt\exp\left({-\mu \Gamma_{3456}\Gamma_{0}\sigma^{0}}\right)T^0\eta.}}

We now show that the Noether charges \generators\ 
of the deformed field theory \newlag\ satisfy the Type IIB plane wave   superalgebra. For the  even generators we get:
\eqn\comev{\eqalign{[P^I,H]=i\mu^2 K^I\qquad[K^I,H]=-iP^I\qquad [P^I,K^J]=i\delta^{IJ}P^+\cr
[P^A,J^{BC}]=-\delta^{AB}P^C+\delta^{AC}P^B\qquad[P^{A'},J^{B'C'}]=-\delta^{A'B'}P^{C'}+\delta^{A'C'}P^{B'}\cr
[K^A,J^{BC}]=-\delta^{AB}K^C+\delta^{AC}K^B\qquad[K^{A'},J^{B'C'}]=-\delta^{A'B'}K^{C'}+\delta^{A'C'}K^{B'}\cr
 \hskip-50pt [J^{AB},J^{CD}]=-\delta^{BC}J^{AD}+\delta^{AC}J^{BD}+\delta^{BD}J^{AC}-\delta^{AD}J^{BC}\cr
 [J^{A'B'},J^{C'D'}]=-\delta^{B'C'}J^{A'D'}+\delta^{A'C'}J^{B'D'}+\delta^{B'D'}J^{A'C'}-\delta^{A'D'}J^{B'C'}.}}
The commutation relations between odd and even generators are:
\eqn\comodev{\eqalign{[P^I,Q]=-i\mu\Gamma^I\Gamma_{3456}q\qquad[K^I,Q]=-i\Gamma^I\Gamma^0 q\cr
[H,Q]=0\qquad
[H,q]=-i\Gamma_{3456}\Gamma^0q\cr
[J^{AB},Q]=- {1\over 2} \Gamma^{AB}Q\qquad[J^{AB},q]= -{1\over 2} \Gamma^{AB}q\cr
[J^{A'B'},Q]=- {1\over 2} \Gamma^{A'B'}Q\qquad[J^{A'B'},q]= -{1\over 2}\Gamma^{A'B'}q.}}
The anticommutators of the supercharges are:
\eqn\anticom{\eqalign{\{q^\alpha,q^\beta\}=i\delta^{\alpha\beta}P^+\qquad\{q^\alpha,Q^\beta\}=-{i\over 2}(\Gamma^I\Gamma^0)^{\alpha\beta}P^I-\mu {i\over 2}(\Gamma_{3456}\Gamma^I)^{\alpha\beta} K^I\cr
\{Q^\alpha,Q^\beta\}=2H\delta^{\alpha\beta}+i\mu(\Gamma^{AB}\Gamma_{3456}\Gamma^0)^{\alpha\beta}J^{AB}+i\mu(\Gamma^{A'B'}\Gamma_{789(10)}\Gamma^0)^{\alpha\beta}J^{A'B'}.}}
This is the superalgebra of the Type IIB plane wave \BlauNE (see also 
\lref\MetsaevRE{
  R.~R.~Metsaev and A.~A.~Tseytlin,
  ``Exactly solvable model of superstring in plane wave Ramond-Ramond
  background,''
  Phys.\ Rev.\  D {\bf 65}, 126004 (2002)
  [arXiv:hep-th/0202109].
}
\MetsaevRE\
for a useful summary of the superalgebra)

\listrefs

\end

 new for nserc \Gomispp